\documentclass[prl,twocolumn,showpacs]{revtex4}

\usepackage{graphicx}
\usepackage{amsmath}
\usepackage{epsfig}
\usepackage{epstopdf}

\def\begeq{\begin{equation}}
\def\endeq{\end{equation}}

\def\begeqar{\begin{eqnarray}}
\def\endeqar{\end{eqnarray}}

\def\aeq{\stackrel{_{\rm a}}{=}}
\def\aapp{\stackrel{_{\rm a}}{\approx}}

\begin{document}

\title{Exact valence bond entanglement entropy and probability distribution \\
in the XXX spin chain and the Potts model}

\author{J.L. Jacobsen$^{1,2}$, H.~Saleur$^{2,3}$}
\affiliation{${}^1$LPTMS, Universit\'e Paris-Sud, B\^atiment 100, 
91405 Orsay, France}
\affiliation{${}^2$Service de Physique Th\'eorique, CEN Saclay,
91191 Gif Sur Yvette, France}
\affiliation{${}^3$Department of Physics,
University of Southern California, Los Angeles, CA 90089-0484}

\date{\today}

\begin{abstract}
  By relating the ground state of Temperley-Lieb hamiltonians to
  partition functions of 2D statistical mechanics systems on a half
  plane, and using a boundary Coulomb gas formalism, we obtain in
  closed form the valence bond entanglement entropy as well as the
  valence bond probability distribution in these ground states. We
  find in particular that for the XXX spin chain, the number $N_{\rm
    c}$ of valence bonds connecting a subsystem of size $L$ to the
  outside goes, in the thermodynamic limit, as $\langle N_{\rm
    c}\rangle (\Omega) = {4\over \pi^2}\ln L$, disproving a recent
  conjecture that this should be related with the von Neumann entropy,
  and thus equal to ${1\over 3\ln 2}\ln L$. Our results generalize to
  the $Q$-state Potts model.
\end{abstract}



\maketitle

\noindent{\bf Introduction.}
Entanglement is a central concept in quantum information processing,
as well as in the study of quantum phase transitions. One of the
widely used entanglement measures is the von Neumann entanglement
entropy $S_{\rm vN}$, which quantifies entanglement of a pure quantum
state in a bipartite system.  To define $S_{\rm vN}$ precisely, let
$\rho = |\Psi\rangle\langle\Psi|$ be the density matrix of the system,
where $|\Psi\rangle$ is a pure quantum state.  Given a complete set
$X$ of commuting observables, let $X = A \cup B$ be a bipartition
thereof. Then $S_{\rm vN}$ is defined as
\begin{equation}
 S_{\rm vN}(A) = - {\rm Tr}_A \, \rho_A \ln \rho_A
\end{equation}
where $\rho_A = {\rm Tr}_B \, \rho$ is the reduced density matrix with
respect to $A$. One readily establishes that $S_{\rm vN}(A) = S_{\rm
  vN}(B)$. In most applications, the subset $A$ corresponds to a
commuting set of observables characterizing only a part of the whole
system, and so 
one may think of $A$ denoting simply that subsystem.

Critical ground states in 1D are known to have entanglement entropy
that diverges logarithmically in the subsystem size with a universal
coefficient proportional to the central charge $c$ of the associated
conformal field theory. Let $L = |A|$ be the size of the subsystem,
and $N$ the size of the whole system, both measured in units of the
lattice spacing, with $1 \ll L \ll N$ (as we shall invariably assume
in what follows). Then \cite{Holzhey,CardyCalabrese}
\begin{equation}
 S_{\rm vN}(A) \aeq (c/3) \, \ln L \,.
 \label{SvNc}
\end{equation}
where by $\aeq$ we denote asymptotic behaviour. Away from the critical point, the entanglement entropy saturates to a
finite value, which is related to the correlation length.

The von Neumann entanglement entropy is not an easy quantity to
calculate, analytically or numerically; nor is it easy to grasp
intuitively. The plethora of algebraic and geometric reformulations of
quantum hamiltonians in one and higher dimensions
\cite{Martinbook,MartinSaleur} suggests that there should be a more
convenient, geometric way to define an entanglement entropy.  In the
case of systems admitting infinite randomness fixed points, in one
\cite{MooreRefaela,MooreRefaelb,Santachiara} as well as in higher
\cite{Igloi,Haas} dimensions, the ground state $|\Omega\rangle$ can be
represented as a single valence bond state, and $S_{\rm vN}$ coincides
with the number of singlets that cross the boundary of
$A$ times the logarithm of the number of states per site.

In an interesting recent paper \cite{Alet} (see also \cite{otherprl} for a related, independent work) 
it was suggested that, even when
$|\Omega\rangle$ is not a single valence bond state but a superposition of
such states, the average number of singlets $N_{\rm c}(\Omega)$ crossing the
boundary of the subsystem (multiplied e.g.~by $\ln 2$ for spins $1/2$) could
still be used as a measure of the entanglement entropy with all the required
qualitative properties. Moreover, it was observed numerically in \cite{Alet}
that, up to statistical errors, this valence bond entanglement entropy $S_{\rm
  VB}$ had the same asymptotic behavior as $S_{\rm vN}$ for the XXX quantum
spin chain, namely (\ref{SvNc}) with $c=1$. In other words, the observation of
\cite{Alet} was that
\begin{equation}
 \langle N_{\rm c}\rangle (\Omega)\aapp {1\over 3\log 2} \, \ln L
 \simeq 0.481 \ln L \,.
\label{obsAlet}
\end{equation}
Apart from this, the valence bond basis has been actively studied
recently, in particular from rigorous \cite{Mambrini,BeachSandvik} and
probabilistic \cite{Sandvik} points of view.
 
Studying entanglement entropy from a geometrical point of view seems
particularly appealing in view of the quantum dimer model of Rokhsar
and Kivelson \cite{RokhsarKivelson} and its many recent
generalizations. Indeed, Henley \cite{Henley} has shown that for {\em
  any} classical statistical mechanics model equipped with a discrete
state space and a dynamics satisfying detailed balance, there is a
corresponding quantum Hamiltonian whose ground state $|\Omega\rangle$
is precisely the classical partition function. Of particular interest
are then statistical mechanics models in which the microscopic degrees
of freedom directly {\em define} the valence bonds. This is the case
for a certain class of lattice models of loops, to be studied below.

We show in this Letter that the probability distribution of the number
of singlets crossing the boundary can be exactly determined for the
XXX spin chain as well as for the related $Q$-state Potts model
hamiltonians. We find that (\ref{obsAlet}) is not quite correct: the
exact leading asymptotic behaviour is in fact $\langle N_c\rangle
(\Omega) \aeq {4\over \pi^2}\ln L \simeq 0.405\ln L$. All other cumulants
have similar closed form expressions.

\smallskip

\noindent {\bf Entanglement and the TL algebra.} The 2D classical
$Q$-state Potts model can be defined for $Q$ non integer through
an algebraic reformulation where $Q$ enters only as a
parameter. For this, recall that the transfer matrix in the
anisotropic limit gives rise to the hamiltonian \footnote{The scale of
  (\ref{hamil}) affects the sound velocity and is important when
  studying the scaling of gaps. But it does not matter when dealing
  with entanglement issues.}
\begin{equation}
H=-\sum_{i=1}^{N-1} e_i
\label{hamil}
\end{equation}
Here the $e_i$ are elements of an associative unital algebra called
the Temperley-Lieb (TL) algebra, defined by
\begin{eqnarray}
 e_i^2 &=& \sqrt{Q}e_i\nonumber\\
 \left[e_i,e_j\right] &=& 0 \mbox{ for } |i-j|\geq 2 \nonumber\\
 e_ie_{i+1}e_i &=& e_i
\end{eqnarray}

\begin{figure}
\includegraphics[width=6cm,angle=0]{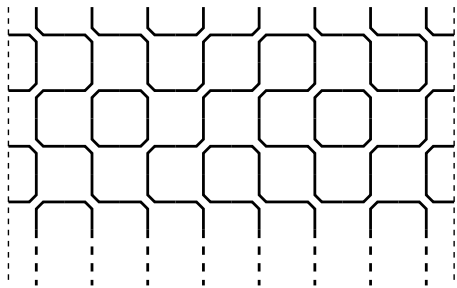}
\caption{Loop representation of the periodic TL algebra (with $N=8$) on the
infinite half cylinder $C_-$. The basis state corresponding to the upper rim
is $\bullet(())()\bullet$.}
\label{fig:conf}
\end{figure}

The ground state of $H$ depends on the representation of the algebra.
For our purpose, it is natural to use the loop model representation,
where the generators act on the following non-orthogonal but linearly
independent basis states. Each basis state corresponds to a pattern of
$N$ parentheses and dots, such as $()\bullet (())\bullet$.  The
parentheses must obey the typographical rules for nesting, and the
dots must not be inside any of the parentheses. These rules imply that
the $()$ pairs consist of one even and one odd site, and that dots are
alternately on even and odd sites. We start by convention with an odd
site on the left. Note that we have used an open chain for
convenience, but that a periodic chain can be considered as well. This
requires the introduction of an additional generator $e_N$ coupling
the $N^{\rm th}$ and first site, and in the graphical representation,
parentheses can now be paired cyclically so $)\bullet (())\bullet ($
is now a valid pattern. This is shown in Fig.~\ref{fig:conf}.

One can interpret these states in terms of spin $1/2$ by
associating with each pair of nearest parentheses $()$ a $U_qsl(2)$
singlet ($\sqrt{Q}=q+q^{-1}$) so a valence bond can be drawn between
the two corresponding sites. (Note that the generator $e_i$ is nothing
but the operator that projects sites $i$ and $i+1$ onto the singlet.)
For the dots, the state must be chosen such that the application of
the projection operator onto the $U_qsl(2)$ singlet for any two dots
that are adjacent (when parentheses are ignored) annihilates the
state.  Thus those sites are ``non-contractible''. With these
definitions, it is clear that the TL algebra does not mix states with
different numbers of non-contractible sites. For $Q$ generic, the set
of basis states with fixed number $2j$ of such sites provides an
irreducible representation of TL, of well-known dimension
\begin{equation}
 d_j={N\choose N/2+j}-{N\choose N/2+j+1}
\end{equation}
where $N/2+j$ must be an integer. This dimension coincides with the
number of representations of spin $j$ appearing in the decomposition
of the product of $N$ spins $1/2$. This is no accident: it is
well-known \cite{Affleck} that the uncrossed diagrams are linearly
independent and form a basis of the spin $j$ sector in the $sl(2)$
case; the results extends trivially to the $U_qsl(2)$ case with $q$
generic.
 
When $\sqrt{Q} \ge 0$ the ground state is found in the sector with
$j=0$ for $N$ even (and $j=1/2$ for $N$ odd).  Note that the
valence-bond basis is not orthonormal.  The simplest way to proceed is
thus not to calculate matrix elements of the hamiltonian $H$ in this
basis $\langle w_i|H|w_j\rangle$ but rather to define a non-symmetric
matrix $h_{ij}$ by expressing the action of $H$ on any state as a
linear combination of states
\begin{equation}
 H|w_i\rangle=\sum_j h_{ij}|w_j\rangle
\end{equation}
The matrix $h_{ij}$ is unique due to the linear independence of the
states. The eigenvalues and right eigenvectors of $h$ give those of
$H$.
 
Since all entries $h_{ij}$ are (strictly) positive, the
Perron-Frobenius theorem implies that the ground state
$|\Omega\rangle$ expands on the basis words with positive
coefficients \footnote{Note that the scalar product of a state
  $|w\rangle$ with itself is equal to $\sqrt{Q}^{N/2-j}$ for all $w$,
  so there is no need to consider ``normalized'' basis words.}
\begin{equation}
 |\Omega\rangle=\sum \lambda_w |w\rangle, \qquad \lambda_w>0
 \label{positivity}
\end{equation}
We define the number of valence bonds $N_{\rm c}$ connecting the subsystem
to the outside as the number of unpaired
parentheses in the subsystem. We are here interested in its mean value
\begin{equation}
 \langle N_{\rm c}\rangle (\Omega) =
 {\sum_w \lambda_w N_{\rm c}(w)\over \sum_w\lambda_w}
\end{equation}
and more generally in the probability distribution
\begin{equation}
p(N_{\rm c})={\sum_{w: N_{\rm c}(w)=N_{\rm c}} \lambda_w \over \sum_w\lambda_w}
\label{distribution}
\end{equation}
Below we establish the leading asymptotic behaviour of $\langle N_{\rm
  c}\rangle $ (and the higher cumulants) in the scaling limit $1 \ll L
\ll N$. Note that the TL formulation shows relationship between the
Potts hamiltonian when $\sqrt{Q}=2\cos{\pi\over k+2}$, with $k$ integer,
and the interacting anyons (coming in $k+1$ species) hamiltonian in
\cite{Feiguin}. The valence bond entanglement entropy can be defined
for these models as well, and, in the sector of vanishing topological
charge, coincides with the one we are studying.

\smallskip

\noindent{\bf Mapping onto a boundary problem.}
The wave function in the ground state of a certain hamiltonian with
periodic (free) boundary conditions \footnote{The boundary conditions
  at large $N$ are not expected to affect the leading behaviour of
  $S_{\rm VB}$.} can be obtained as the path integral of
the equivalent Euclidian theory on a infinite half cylinder (annulus),
denoted $C_-$ (or $A_-$). To translate this in
statistical mechanics terms, note that if we consider the square
lattice with axial (diagonal) direction of propagation
[cf.~Fig.~\ref{fig:conf}], the hamiltonian belongs to a family of
commuting transfer matrices describing the $Q$-state Potts model with
various degrees of anisotropy.  The ground state of all these transfer
matrices is given by $|\Omega\rangle$. Let us chose for instance the
particular case where the Potts model is isotropic, with coupling
constant $e^K=1+\sqrt{Q}$.  Now the ground state $|\Omega\rangle$ can
be obtained by applying a large number of times the transfer matrix on
an arbitrary initial state (corresponding to boundary
conditions at the far end of $C_-$ or $A_-$.)  Clearly, by the mere
definition of the transfer matrix, this means that the coefficients of
the ground state $|\Omega\rangle$ on the words $|w\rangle$ are (up to
a common proportionality factor) equal to the partition function of
the 2D statistical system on $C_-$ or $A_-$ with boundary conditions
specified by $|w\rangle$.

We must now study such partition functions. We move
immediately to the limit $N\rightarrow \infty$. We then have a system
in the half plane, which, in the geometrical description, corresponds
to a gas of loops with fugacity $\sqrt{Q}$ in the bulk, with half
loops ending up with open extremities on the boundary. To go to the
continuum limit it is convenient to transform this loop model into a
solid-on-solid model \cite{Nienhuis}.  For this parametrize
$\sqrt{Q}=2\cos \pi e_0$ with $0 \le e_0 < 1$ \footnote{When $\sqrt{Q}
  < 0$ (i.e.~$\frac12 < e_0 < 1$) the true ground state has spin
  $j>0$. Nevertheless, we continue to let $|\Omega\rangle$ denote the
  $j=0$ ground state. Numerical studies then indicate that
  (\ref{positivity}) holds even for $\sqrt{Q}<0$, if $N$ is large
  enough.}  Give to all the loops an orientation, and introduce
complex weights $\exp\left(\pm {i\pi e_0\over 4}\right)$ for the left
and right turns.  Since on the square lattice the number of left
($n_L$) minus the number of right ($n_R$) turns equals $\pm 4$, this
gives closed loops the correct weight $\sqrt{Q}$. Meanwhile, loops
ending on the boundary will get, with this construction, the weight
$\sqrt{Q}_{\rm b}=2\cos{\pi e_0\over 2}$ since for them $n_L-n_R=\pm
2$ \cite{Ivan}. Although no such boundary weight appeared in the
initial lattice model and partition function, we note that for the
fully packed loop model we are interested in it does not matter: the
number of open loops touching the boundary is just $N/2-j$, a
constant.

Introducing this boundary loop weight allows complete mapping to the
SOS model (or free six-vertex model). Now it is known that in the
continuum limit, the dynamics of the SOS height variables turns into
the one of a free bosonic field \cite{Nienhuis}. In a renormalization
scheme where loops carry a constant height step $\Delta\Phi=\pm \pi$,
the propagator of the field evaluated at two points $x,x'$ on the
boundary reads, in the infinite size limit \cite{Ivan}
\begin{equation}
\langle\Phi(x)\Phi(x')\rangle_{\rm N}=-{1\over g}\ln\left|x-x'\right|^2
\end{equation}
where $g=1-e_0$. Here the subscript ${\rm N}$ indicates Neumann
boundary conditions, corresponding to the presence of loop extremities
on the boundary.

Let us now single out a segment of length $L$ on this boundary and
attempt to count the number of loops connecting this segment to the
rest of the boundary. To do this we insert a pair of vertex operators,
one at each a extremity of the segment, $V=\exp\left[i(\pm
  e_1+e_0/2)\Phi\right]$.  These operators do not affect the loops
encircling the whole interval $L$ since they modify the weight of such
loops from $e^{\pm i\pi e_0/2}$ to $e^{\pm i\pi e_0/2}e^{\mp i\pi
  e_0}=e^{\mp i\pi e_0/2}$, thus giving the same sum $\sqrt{Q}_{\rm
  b}$. But for loops connecting the inside to the outside, the weight
is now $w=2\cos\pi e_1$. The boundary dimension of the fields $V$ is,
using the propagator
\begin{equation}
h={4e_1^2-e_0^2\over 4g}
\end{equation}
so their two-point function decays as $L^{-2h}$. We can then find the
average number of loops separating two given points by taking
appropriate derivatives, and setting $e_1=e_0/2$ in the end.
This leads to our main result
\begin{equation}
 \langle N_{\rm c}\rangle(\Omega) \aeq
 {e_0\over \pi (1-e_0)} {2\cos(\pi e_0/2)\over \sin (\pi e_0/2)}\ln L
\label{main}
\end{equation}
For the XXX chain ($e_0=0$) this reads $\langle N_{\rm
  c}\rangle \aeq {4\over \pi^2}\ln L\approx 0.405\ln L$, while for bond
percolation ($Q=1$ or $e_0=1/3$) we have $\langle N_{\rm c}\rangle \aeq
{\sqrt{3}\over \pi} \ln L \approx 0.551\ln L$.  The slope becomes $1$
exactly as $e_0\rightarrow 1$, or $\sqrt{Q}\rightarrow -2$.
We note that the result for the XXX case is close but definitely
different from the one proposed in \cite{Alet}. 

It is amusing to observe that one can exactly interpret the valence
bond as singlet contractions for an ordinary supergroup in the case
$Q=1$, by taking a lattice model where the fundamental
three-dimensional representation of $SU(2/1)$ and its conjugate
alternate. The hamiltonian is again (\ref{hamil}), but this time the
$e_i$ are projectors onto the singlet in $3\otimes \bar{3}$. The
effective central charge for the this spin chain is $c_{\rm
  eff}=1+{9\over \pi^2} [\hbox{arccosh}(3/2)]^2 \simeq 1.845$, and
extending the argument suggested in \cite{Alet} for the XXX case gives
a slope of ${c_{\rm eff}\over 3\ln 3} \simeq 0.559$, even closer to the
exact result (\ref{main}).

Of course, by taking higher derivatives of the two-point function of
the vertex operators one can access the higher moments of (\ref{distribution}).
In fact, the two-point function
itself is nothing but the characteristic function of $p(N_{\rm c})$,
although carrying out the Fourier transform in general is somewhat
cumbersome. We will content ourselves here by giving the first few
cumulants ${\cal C}_k = (c_k/\pi^k) \ln L$, with, in the XXX case (top) and
the $Q=1$ case (bottom):
\begin{equation}
 c_1 = \left \lbrace \begin{array}{ll} 4 \\ \sqrt{3}\pi
 \end{array} \right. \ 
 c_2 = \left \lbrace \begin{array}{ll} 8/3 \\ 2(2\pi\sqrt{3}-9)
 \end{array} \right. \ 
 c_3 = \left \lbrace \begin{array}{ll} 16/15 \\ 8(5\pi\sqrt{3}-27)
 \end{array} \right.
\end{equation}
together with the observation that, as $\sqrt{Q}\rightarrow -2$, the
probability distribution becomes Poissonian:
\begin{equation}
 \lim_{\sqrt{Q}\rightarrow -2} P(N_{\rm c}) = 
 e^{-\ln L}{(\ln L)^{N_{\rm c}}\over N_{\rm c}!}
\end{equation}

\smallskip

\begin{figure}
\vskip-1.0cm
\includegraphics[width=7cm,angle=270]{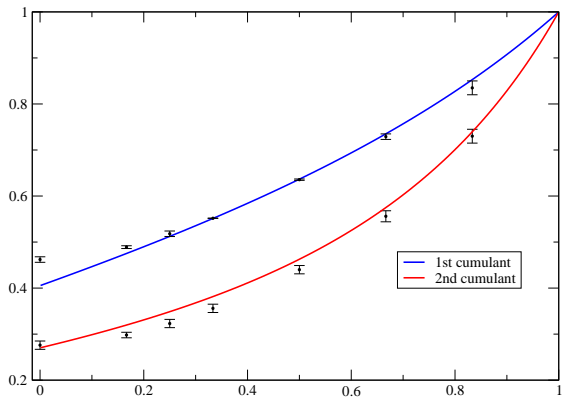}
\vskip-1.0cm
\caption{Comparison between exact and numerically determined values of the
slopes $c_1$ and $c_2$, shown as functions of the parameter $e_0$.}
\label{fig:cums}
\end{figure}

\noindent{\bf Numerical calculations.} We have computed the
distribution (\ref{distribution}) numerically by exactly diagonalizing
the transfer matrix, for periodic chains of size up to $N_{\rm max}=32$. The
cumulants ${\cal C}_k \propto \ln L$ of $p(N_{\rm c})$ obey a very simple
finite size scaling (FSS) form, where $\ln L$ has to be replaced by ${N\over
  \pi}\ln \sin{L\pi\over N}$; this follows from standard formulas for
two-point functions of our vertex operators $V$. Precise values of the
slopes $c_k$ can then be extracted from a careful analysis of the residual 
FSS effects. As shown in Fig.~\ref{fig:cums} they agree well with our
analytical results, except for $Q \to 4$, where we expect logarithmic
FSS corrections.

For $Q=1$, the combinatorial nature of $|\Omega\rangle$ implies that
all $\lambda_w$ in (\ref{positivity}) are {\em integers}. This allows
to obtain $p(N_{\rm c})$ {\em exactly} for finite $L$ and $N \le
N_{\rm max}$.  Using this, we can in some cases conjecture $p(N_{\rm
  c})$ for any value of $N$ \cite{combinatorics}. In particular we
have established that
\begin{equation}
 \langle N_{\rm c}\rangle = (N^2-L^2) p_k(N^2)
 \prod_{n=0}^{N/2-1} (N^2-(2n+1)^2)^{n-N/2}
\end{equation}
where $p_k$ is a polynomial of degree $k=\frac18 (N+4)(N-2)$ in
$N^2$. This exact FSS form allows to obtain for the slope $c_1 =
0.5517 \pm 0.0003$, in very precise agreement with the value
$0.551329$ from (\ref{main}).

\smallskip

\noindent{\bf Conclusions.} As already argued in \cite{Alet}, $S_{\rm
  VB}$ is a measure of entanglement which seems as good qualitatively
as $S_{\rm vN}$, and easier to obtain numerically.  We have shown in
this Letter that it is also possible to tackle it analytically in the
1D case. The results are less easily expressed than for $S_{\rm vN}$
(which is proportional to $c$) . On the other hand, they fit
considerably more naturally within the transfer matrix and Coulomb gas
formalism. It remains to be seen what happens for other models, and
whether in particular a c-theorem of sorts is obeyed for $S_{\rm VB}$.

\smallskip

\noindent{\bf Acknowledgments.} We thank F. Alet for helpful
exchanges, and for pointing out Ref.~\cite{Feiguin}.  This
work was supported by the European Community 
Network ENRAGE
(grant MRTN-CT-2004-005616) and by the Agence Nationale de la
Recherche (grant ANR-06-BLAN-0124-03).

\end{document}